\documentclass[twocolumn,prb,showpacs]{revtex4}
\usepackage{graphicx}
\pdfoutput=1

\begin{document}
\title{Spin-1 Heisenberg antiferromagnetic chain with exchange and
  single-ion anisotropies}
  \author{D.~Peters}
  \affiliation{Institut f\"ur Theoretische Physik,
    RWTH Aachen,
    52056 Aachen, Germany}
  \author{I.~P.~McCulloch}
  \affiliation{Department of Physics, University of Queensland, 
    Brisbane, QLD 4072, Australia}
  \author{W.~Selke}
  \affiliation{Institut f\"ur Theoretische Physik,
    RWTH Aachen, and JARA-SIM, 52056 Aachen, Germany}

\begin{abstract}

Using density matrix renormalization group calculations, ground state
properties of the spin-1 Heisenberg chain with exchange and
single--ion anisotropies in an external field are studied. Our
findings confirm and refine recent results by Sengupta and
Batista, Physical Review Letters 99, 217205 (2007), on the same
model applying Monte Carlo techniques. In particular, we present
evidence for two types of biconical (or supersolid) and for
two types of spin--flop (or superfluid) structures. Basic
features of the quantum phase diagram may be interpreted qualitatively in
the framework of classical spin models.
\end{abstract}


\maketitle

Recently, low-dimensional quantum anisotropic Heisenberg 
antiferromagnets have been shown to exhibit
the analogue of the supersolid phase \cite{sen,flora,picon}, usually
denoted in magnetism as 'intermediate', 'mixed' or biconical \cite{koster}
phase, in which both order parameters of the bordering antiferromagnetic
and spin--flop phases do not vanish. Indeed, already some
decades ago, in 1956, Matsubara and Matsuda \cite{matsu} pointed out
the correspondence between quantum lattices and anisotropic
Heisenberg models, when expressing
Bose operators by spin operators. Using mean-field theory
for calculating ground--state
and thermal properties, supersolid or biconical structures have been  
observed in the uniaxially anisotropic XXZ Heisenberg antiferromagnets with 
additional single--site terms due to crystal--field anisotropies 
or with more--than--nearest neighbor interactions \cite{tsu,liu} (note
that the mean-field approximation of the quantum models corresponds to that of
classical models). Such
phases may give rise to interesting multicritical behavior, 
especially, to tetracritical points \cite{aha,folk}. 

In the last few years, biconical structures and
phases in classical
XXZ Heisenberg antiferromagnets with and without single--ion
anisotropies in two as well as three dimensions have been
studied using ground state considerations and
Monte Carlo techniques \cite{hws,hs,bs}.

Experimental evidence for biconical phases has been accumulated 
over the years \cite{bast,smee,zhou,ng,bog}.

The current search for biconical phases in quantum
magnets \cite{flora,sen,picon} seems
to be partly motivated by the fact that 
they are analogues to the supersolid phases \cite{wes,schmidt,nuss}. Of
course, it is also of much interest to study the impact of quantum
fluctuations on the phases known to occur
in classical anisotropic Heisenberg antiferromagnets in a
magnetic field. In the following Note we shall address
both aspects. 

Specifically, we shall analyze ground state properties, $T=0$, of
the spin--1 XXZ Heisenberg
antiferromagnetic chain with a single--ion
anisotropy in a field $B$. Using quantum Monte
Carlo simulations, namely
stochastic series expansions, for chains with periodic boundary
conditions, Sengupta and Batista
showed that its quantum phase diagram at
zero temperature displays a
field--induced supersolid phase \cite{sen}. The model is
described by the Hamiltonian   

\begin{eqnarray}
{\cal H} &=& \sum\limits_{i} 
   (J (S_i^x S_{i+1}^x + S_i^y S_{i+1}^y
    + \Delta S_i^z S_{i+1}^z)\nonumber\\
   \label{Ham}
   & &
   {}+ D (S_i^z)^2  - B S_i^z) 
\end{eqnarray}

\noindent
where $i$ denotes the lattice sites. For $\Delta>1$ and $D>0$, the
exchange and single--ion terms describe competing, uniaxial (along the
direction of the field $B$, $B>0$, the $z$--direction) and
planar anisotropies. Following the previous analysis \cite{sen}, we
shall deal with the case $D= \Delta/2$, restricting the analysis
to the $(\Delta,B/J)$--plane.

To study the model, we here apply density
matrix renormalization group (DMRG)
techniques, yielding accurate results at zero
temperature \cite{white,scholl}. In particular, we considered
chains with open boundary conditions (allowing the study of fairly
long chains). To monitor finite--size effects, the
number of sites $L$ ranged from 15 up to 128. Usually, a random state
was chosen as initial state. To get reliable data, typically up
to 500 states during 120 sweeps were kept, with a total truncation error 
of $10^{-7}$ and a total energy variance of about $10^{-5}$.  

At given chain length $L$, exchange anisotropy $\Delta$, and field $B/J$, the
total magnetization $M= \sum\limits_{i}<S_i^z>$ follows from
minimization of the energy $E(M,B)= E_0(M)-MB$,
where $E_0(M)$ is the ground state energy at $B=0$ obtained from 
the DMRG calculations. Here and in the following, brackets, $<...>$,
denote quantum mechanical expectation values. For fixed total
magnetization various physical quantities of interest were
determined, including the profile of the $z$--component
of the magnetization, $m_i= <S_i^z>$, longitudinal and transverse
correlators, $<S_i^zS_{i+r}^z>$ 
and, e.g., $<S_i^xS_{i+r}^x>$, as well as possible
order parameters of the various structures and phases \cite{sen}.

\begin{figure}[h]\centering
        \includegraphics[height=250pt,angle=270]{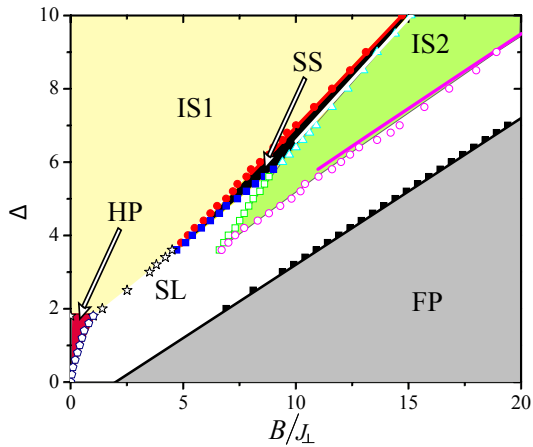}
\caption{\label{fig1} Ground state phase diagram of Hamiltonian (1)
  with $D= \Delta/2$, as has been obtained by
  Sengupta and Batista\cite{sen}. $J_{\perp}$ in this figure 
  is denoted by $J$ in the text.}
\end{figure}

According to the previous analysis \cite{sen}, there are six
distinct phases in the $(\Delta,B/J)$-- plane at $T=0$, as
depicted in Fig. 1. The Haldane phase (HP) \cite{hald} occurs
at small fields and anisotropies. Of course, the ferromagnetic (or
normal fluid in quantum lattices) phase (FP) occurs for all
anisotropies at sufficiently large
fields. Furthermore, there are two solid \cite{tsu} or Ising--like phases: The
usual antiferromagnetic or Neel phase, the 'IS1'--phase in
the notation of Sengupta and Batista, and, at rather
large anisotropies, the 'IS2'--phase with $M \approx L/2$. Actually,
the IS2--phase corresponds to the (10)--phase in the Ising--limit of
Hamiltonian (1), the antiferromagnetic Blume--Capel
model \cite{wang}. The other two ground state phases are
the superfluid \cite{tsu} or spin--liquid \cite{sen} (SL) phase, usually called
in magnetism the spin--flop
phase, and the supersolid \cite{tsu} (SS) or biconical \cite{koster} phase.

Indeed our analysis at selected values
of $\Delta$, $3 \le \Delta \le 7$, confirms the phase diagram
\cite{sen}, see Fig. 1. Moreover, we find evidence
for two types of biconical as well as two types of
spin--liquid structures. The evidence stems, especially, from
the magnetization profiles,
$m_i$. These profiles strongly depend on whether the number
of sites, $L$, is odd or even, due to the open boundary
conditions. In addition, finite--size effects may be important, when
attempting to identify the ground state structures and 
transition points of the infinite chain.

\begin{figure}[h]\centering
        \includegraphics[width=250pt]{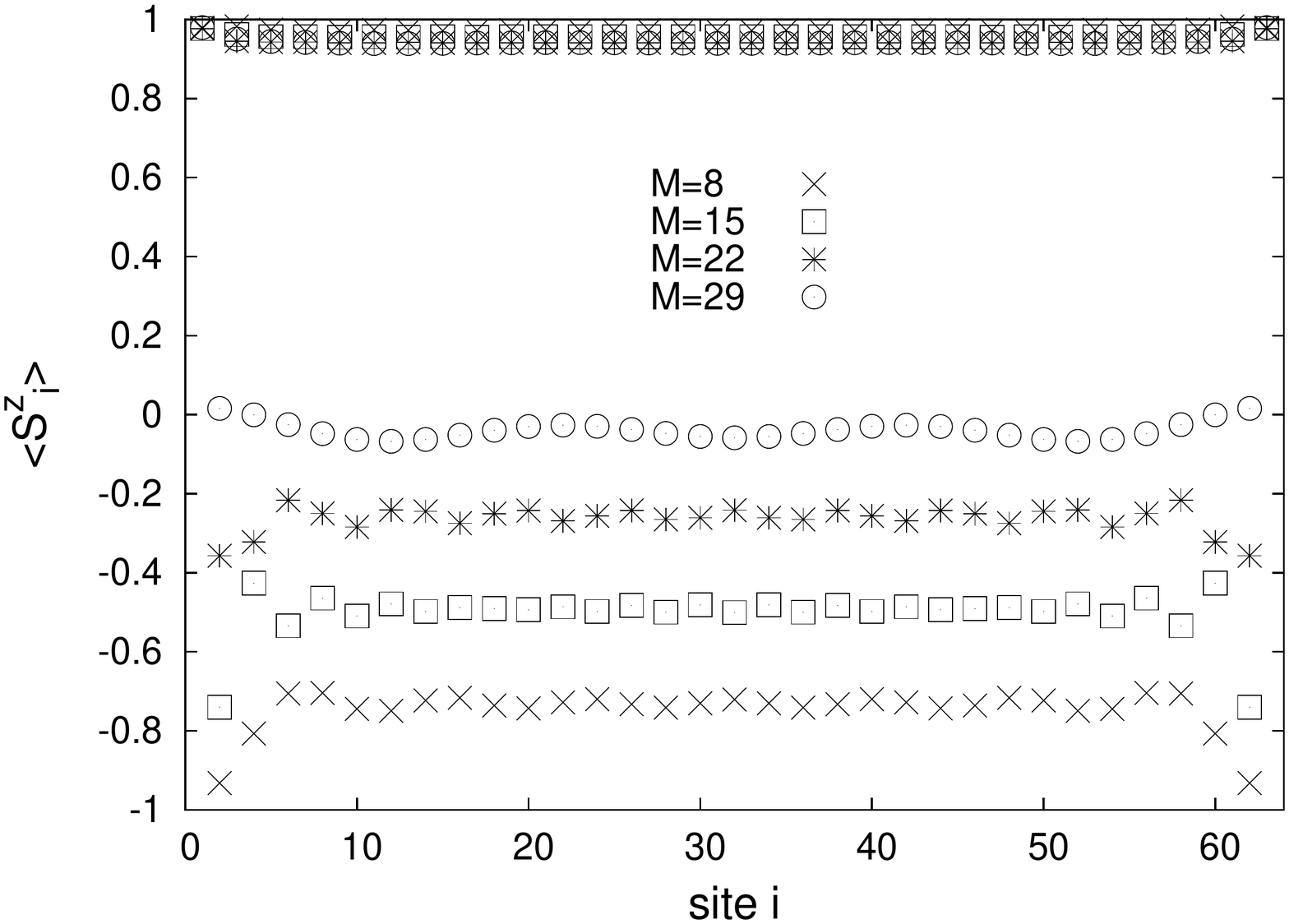}
\caption{\label{fig2}Magnetization profiles $m_i$ for various total
 magnetizations $M$ in the SS phase in between the IS1 and IS2
 phases at $\Delta$= 7 ($L= 63$). The
 lower magnetizations belong to odd sites $i$, the upper ones to
 even sites. The
 profiles describe ground states in the range $10.1 < B/J < 10.7$}.
\end{figure}

As illustrated in Figs. 2 and 3, for odd $L$, the biconical structures
in between the IS1 and IS2 phases differ remarkably from those in between the
IS1 and spin--flop phases. The first situation is exemplified in
Fig. 2 for $\Delta= 7$. We show magnetization profiles $m_i$ in
the supersolid phase. Close to the 
center of the chain, i.e. in the 'bulk', finite effects have
been found to be weak. Increasing the total magnetization $M$ from 0
to about $L/2$ one encounters 
antiferromagnetic and supersolid ground states belonging to
increasing fields $B/J$. In the bulk, $m_i$ is observed
to stay close to one for odd sites $i$, while it changes to almost
zero at the even sites on approach to the IS2 phase, $M \approx
L/2$. One may describe that behavior in two ways: The spins on the
odd sites always point in the field, or $z$, direction, while they
turn, starting from --$z$ direction, in the IS1 phase more and
more with larger values of $M$ towards the $xy$-plane on
the even sites. Alternatively, one may
interpret the behavior in the framework of the antiferromagnetic
Blume--Capel chain. At $T=0$, there is a direct transition from
the antiferromagnetic to the (10)--phase, fixing $D$ and
enlargening the field. Along that transition line, there is
a high degeneracy in configurations, where arbitrary fractions
of spins in the state '--1' of the antiferromagnetic configuration
are replaced by spins in the state '0'. These degenerate structures
seem to give rise to the biconical phase in the strongly 
anisotropic quantum Heisenberg model. 

\begin{figure}\centering
        \includegraphics[width=250pt]{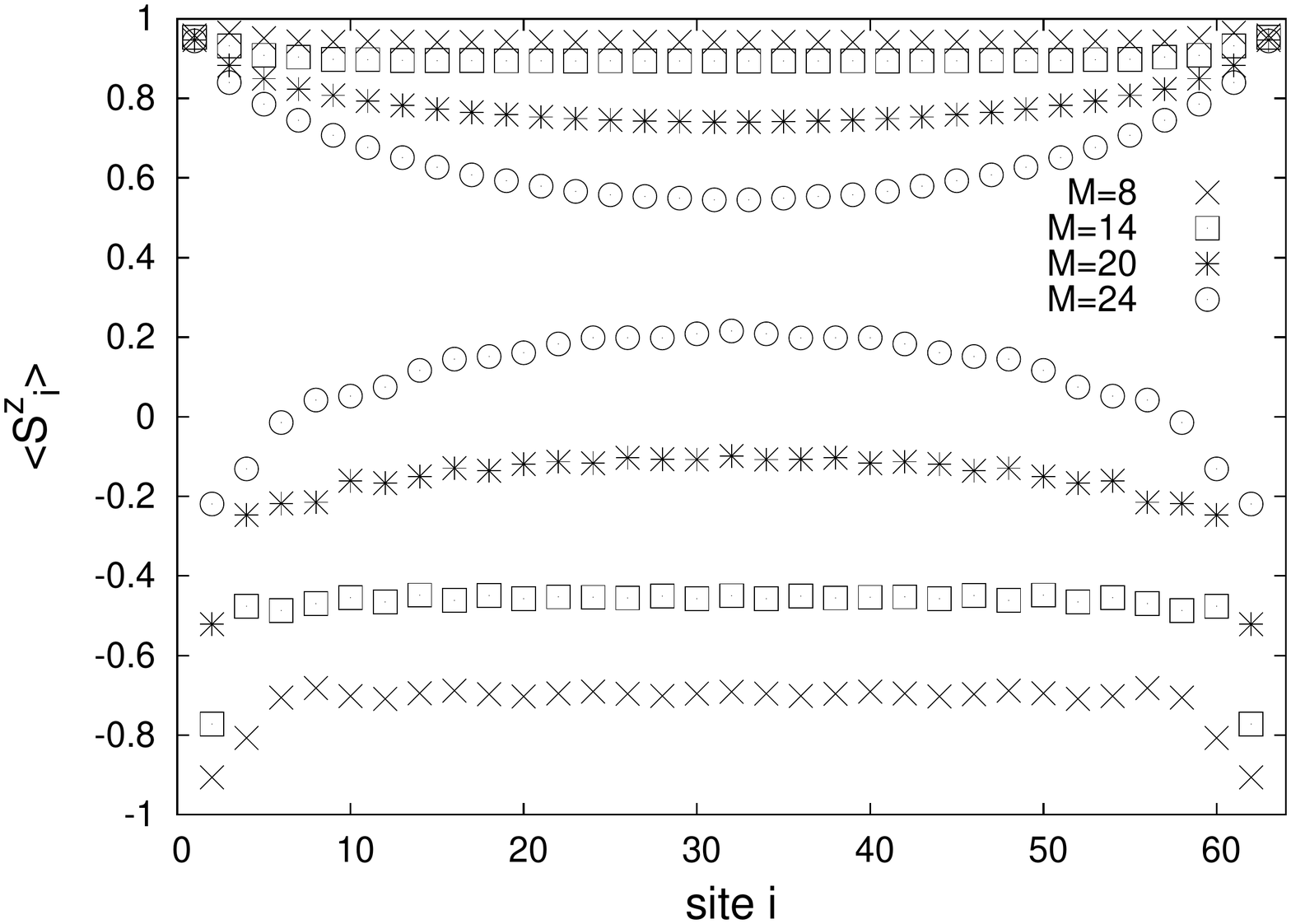}
\caption{\label{fig3}Magnetization profiles in the SS phase in
  between the IS1 and SL phases for various values of the total
  magnetization $M$, at $\Delta= 5$ ($L= 63$). The lower and upper
  values belong to odd and even sites, resp. The profiles describe
  ground states in the range $7.0 <B/J< 7.5$.}.
\end{figure}

Attention may be drawn to the weak, but clearly visible modulations
in the magnetization profiles in these supersolid structures, see Fig. 2.

Let us now consider the supersolid or biconical structures in between
the IS1 and SL phases. Results for local magnetizations, $m_i$, are
depicted in Fig. 3 for $\Delta= 5$. Now, on approach
to the SL phase, the magnetization, in the bulk, at the odd sites
becomes much lower than one, and the difference between the
local magnetizations $m_i$ in the center of the
chain on odd and even sites is getting smaller and smaller. Such
a behavior may be expected from classical anisotropic Heisenberg
antiferromagnets \cite{tsu,liu,hs}. Indeed, for
the classical variant of Hamiltonian (1), the $z$--component of the local
magnetization has been observed for biconical structures in between 
the antiferromagnetic and spin--flop phases to display 
a behavior similar to that found and seen here \cite {hs}. The
spins on both sublattices turn gradually
and intercorrelated towards their common spin--flop orientation.

In contrast to the classical variant of the model, the transverse
components of the spins show no long--range (antiferromagnetic)
order in the quantum case. Instead, the transverse
correlations seem to decay algebraically.

Of course, it would be interesting to clarify whether the two types
of supersolid or biconical structures are separated by a
sharp transition and to locate and charactarize that possible
transition. Our preliminary results suggest that the SL phase seems
to disappear at about $5 < \Delta < 5.5$, see also Fig 1. Note
that in this region there are strong finite--size effects for
the magnetization close to the center of the chain, and care 
is needed to discriminate SS and SL structures. For
instance, at $\Delta=5.0$, we analyzed chains with up to 127
sites, identifying then the spin--flop phase. Actually, the
possible transition between the two distinct
supersolid structures may be argued to occur in that part of
the phase diagram. Detailed investigations would require very long
chains, and they are beyond the scope of this Note.     

In the spin--liquid phase we observe different magnetization pattern
for $M < L/2$ and for $M > L/2$, at all values of $\Delta$ we
studied, i.e. independently of the SL phase being separated
by the IS2 phase or not, see Fig. 1. In Figs. 4 and 5, magnetization
profiles, $m_i$, illustrate both situations.     
   
As exemplified in Fig.4, for $M < L/2$, the local magnetization
displays an extended plateau in the center of the chain in the 
SL phase, similar to the behavior in the spin--flop
phase for classical XXZ Heisenberg
antiferromagnets without and with additional (competing) single--ion
anisotropy \cite{hs}. In Fig. 4 tiny modulations associated with odd and
even sites are seen, which, however, are even reduced when considering
longer chains, possibly, vanishing for infinite chains.

\begin{figure}
  \begin{center}
        \includegraphics[width=250pt]{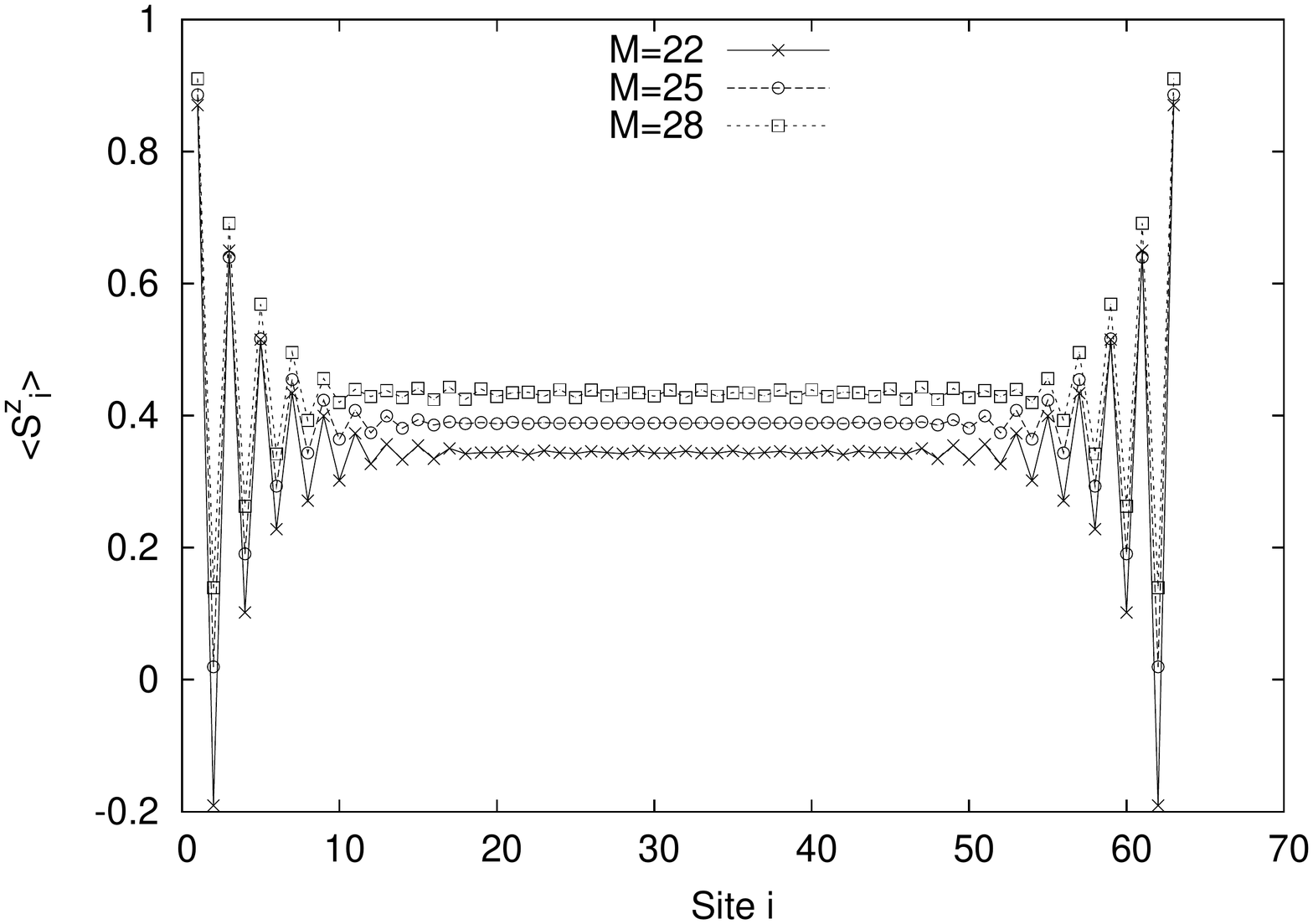}
  \end{center}
  \caption{\label{fig4} Magnetization profiles in the SL phase
    with $M < L/2$ ($L= 63$) at $\Delta= 3.5$. The fields of
    the corresponding ground states are in the range $5.0 <B/J< 5.8$.}
\end{figure}

Increasing $M$, $M> L/2$, but staying in the SL
phase, the magnetization pattern
changes significantly, as depicted in Fig. 5. One first observes
a beat--like modulation about the mean magnetization, as being
well known from superimposing two sine waves with slightly
different wavelengths. The wavelength of the envelope of the beat
is decreasing 
when enlargening $M$. Eventually, the modulation about the increasing
mean value takes on a simple (nearly) sinusoidal form, with the
wavelength getting larger and the amplitude getting smaller
when increasing the total magnetization $M$. Of course, at $M= L$, the
perfect ferromagnetic profile, with $M_i= 1$ for all sites $i$, is 
reached. Going from the beat--like to the sinusoidal profiles, one
increases systematically the average distance between successive
extrema, reflecting, presumably, a quasi-continuous increase of
the winding number of the modulation. Thence, it seems tempting to
suggest the SL phase at $M> L/2$ to be of incommensurate type, while
it seems to be of commensurate type for $M < L/2$.  A similar
distinction for the spin--flop phase may
have been observed before in  different parts of the phase
diagram of Hamiltonian (1) \cite{sakai}.
It is worth mentioning that the amplitude of the modulations
shrink somewhat for longer chains, and it would be interesting
to analyze whether such, presumably, Friedel--like oscillations are
still present in the infinite chain. 

As in the case of the biconical phase,
the transverse components
of the magnetization show no long--range order in the SL phase
of the quantum
model \cite{sen,mik,schulz}, in contrast to the situation
in the classical variant of Hamiltonian (1).

Spin--flop phases with beat--like or sinusoidal modulations in
$m_i$ do not occur in the classical variant of the Hamiltonian (1).

Again, a more detailed analysis, considering carefully finite--size
effects, is desirable, but well beyond the scope of our present
investigation.

\begin{figure}
  \begin{center}
        \includegraphics[width=250pt]{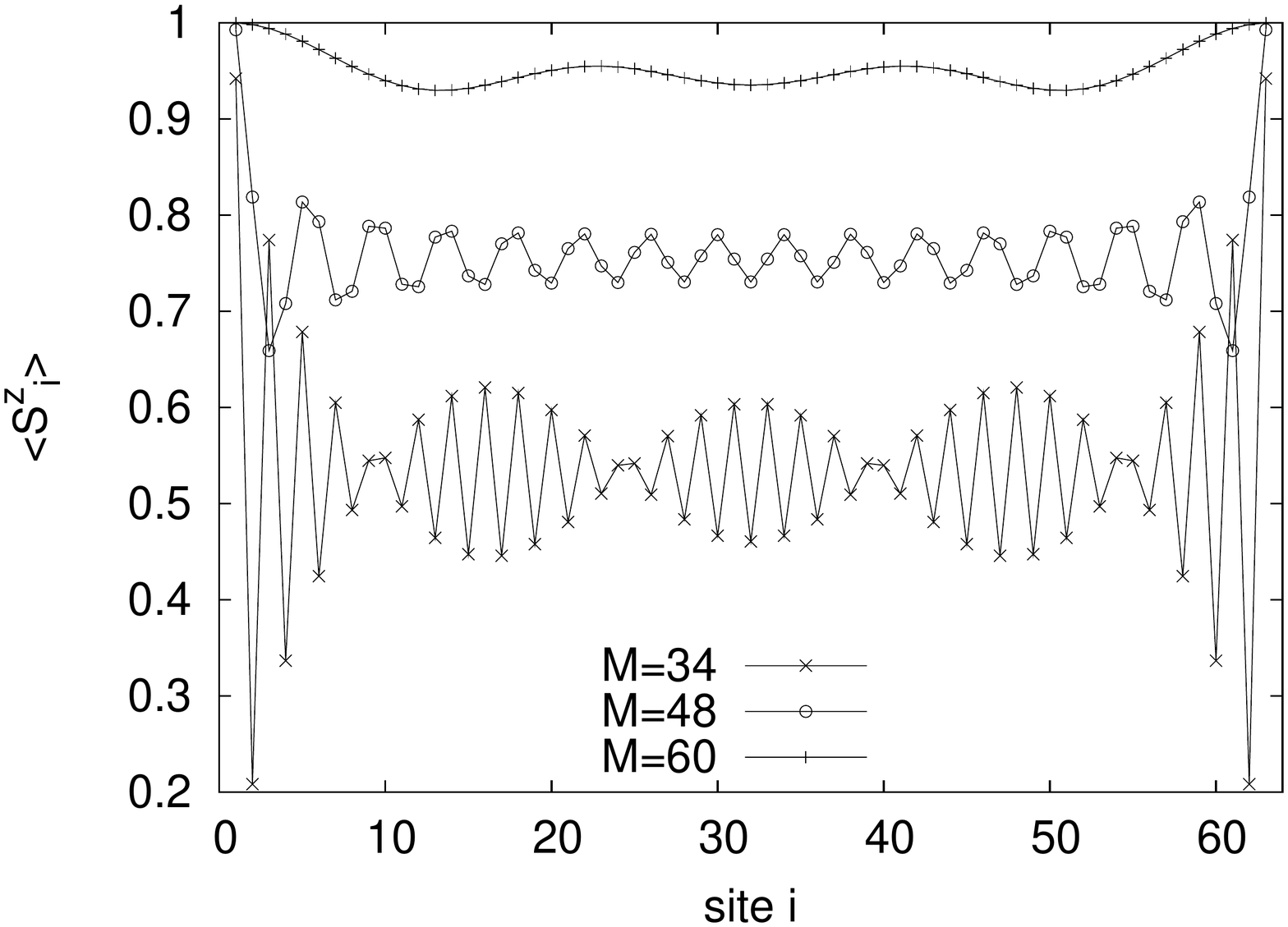}
  \end{center}
  \caption{\label{fig2} Magnetization profiles $m_i$ in the SL phase
    with $M > L/2$ ($L= 63$) at $\Delta= 3.5$. The fields of the 
    corresponding ground states are in the range $6.8 <B/J< 10.7$.}
\end{figure}

To summarize our findings, we have studied the S--1 antiferromagnetic
XXZ Heisenberg antiferromagnetic chain with a competing, planar
single--ion anisotropy. Using DMRG calculations, we confirmed and
refined the ground state phase diagram obtained recently  by Sengupta
and Batista applying Monte Carlo techniques. In particular, we
presented evidence for two
distinct types of biconical or supersolid structures and for
two distinct types of spin--flop or superfluid structures, in the
finite chains we studied. We compared the findings with results on related
classical magnets.  

\acknowledgments
We thank A. Bogdanov, M. Holtschneider, A. Kolezhuk, N. Laflorencie, and
S. Wessel for very useful remarks, discussions, and correspondence, as well as
C.D. Batista and P. Sengupta for sending us figure 1 of this
paper. The research has been funded by the excellence initiative of
the German federal and state governments.

\end{document}